\documentclass[runningheads]{llncs}
\usepackage[T1]{fontenc}
\usepackage{graphicx}%
\usepackage{multirow}%
\usepackage{amsmath,amssymb,amsfonts}%
\usepackage{mathrsfs}%
\usepackage{textcomp}%
\usepackage{manyfoot}%
\usepackage{booktabs}%
\usepackage{listings}%

\PassOptionsToPackage{hyphens}{url}
\PassOptionsToPackage{table}{xcolor}
\usepackage[hidelinks]{hyperref}
\usepackage{tikz}
\usepackage{dsfont}
\usepackage{siunitx}
\usepackage{algorithm2e}
\usepackage{float}

\definecolor{myblack}{RGB}{0, 0, 0}
\definecolor{mygreen}{RGB}{0, 146, 146}
\definecolor{myblue}{RGB}{0, 110, 219}
\definecolor{myred}{RGB}{146, 0, 0}

\begin{document}
\title{Is RISC-V Ready for Machine Learning? Portable Gaussian Processes Using Asynchronous Tasks}
\titlerunning{Portable Gaussian Processes Using Asynchronous Tasks}
%
\author{Alexander Strack\inst{1}\orcidID{0000-0002-9939-9044} \and
	Patrick Diehl\inst{2}\orcidID{0000-0003-3922-8419} \and
	Dirk Pflüger\inst{1}\orcidID{0000-0002-4360-0212}}
\authorrunning{A. Strack et al.}
%

\institute{Institute of Parallel and Distributed Systems, University of Stuttgart,\\ 70569 Stuttgart, Germany\\
	\email{\{alexander.strack, dirk.pflueger\}@ipvs.uni-stuttgart.de} \and
	Applied Computer Science group, Computing and Artificial Intelligence division, Los Alamos National Laboratory, 87544 Los Alamos, NM, USA\\
	\email{diehlpk@lanl.gov}}

\maketitle              
\begin{abstract}

	Gaussian processes are widely used in machine learning domains but remain computationally demanding, limiting their efficient scalability across emerging hardware platforms.
	The GPRat library addresses these challenges using the HPX asynchronous many-task runtime system.
	In this work, we extend GPRat to enable portability across multiple hardware architectures and evaluate its performance on representative x86-64, ARM, and RISC-V chips.
	We conduct node-level strong scaling and problem size scaling benchmarks for Gaussian process prediction and hyperparameter optimization to assess single-core performance, parallel scalability, and architectural efficiency.

	Our results show that while the x86-64 Zen 2 chip achieves a 58\% single-core performance advantage over the ARM-based Fujitsu A64FX, superior parallel scaling allows the 48-core ARM chip to outperform the 64-core Zen 2 by 9\% at full node utilization.
	The evaluated SOPHON SG2042 RISC-V chip exhibits substantially lower performance and weaker scalability, with single-core performance lagging by up to a factor of 14 and large-scale parallel workloads showing slowdowns of up to a factor of 24.
	For problem size scaling, ARM and x86-64 systems demonstrate comparable performance within 23\%.
	These findings highlight the growing competitiveness of purpose-built ARM chips.
	Furthermore, they underscore the importance of wide-register vectorization support and improvements to the memory subsystem for upcoming RISC-V platforms, especially when targeted by many-task runtimes.

	\keywords{Machine Learning \and RISC-V \and ARM \and Gaussian Processes \and Asynchronous Tasks \and HPX \and GPRat}
\end{abstract}
\section{Introduction}\label{sec:introduction}

In recent years, the monopoly of x86-64 architecture chips in high-performance computing (HPC) has been challenged by
emerging reduced instruction set architectures, including ARM~\cite{ARM2021} and RISC-V~\cite{Waterman2014_riscv}.
The first major step for ARM in HPC was marked by Riken's Fugaku supercomputer, based on the Fujitsu A64FX chip, which was the most powerful supercomputer at the time of its deployment.
Regarding RISC-V, the SOPHON SG2042 chip used by the MILK-V Pioneer marks another important step, as it is the first publicly available RISC-V desktop-grade chip suitable for HPC.
As a result of this trend toward greater heterogeneity, software frameworks and scientific libraries must evolve to provide portability across diverse architectures while maintaining performance.

At the same time, Gaussian processes (GPs) have gained substantial traction in the field of machine learning (ML)~\cite{Kocijan2016,Rasmussen2006}.
Unlike neural networks, which rely on a large amount of trainable parameters, GPs follow a non-parametric modeling approach with only a couple of trainable hyperparameters.
Furthermore, GPs naturally provide uncertainty estimates alongside predictions, an essential feature for safety-critical applications.
However, a major limitation of GP methods is their high computational complexity, which poses challenges for large-scale workloads.

To address this challenge, GPRat~\cite{Helmann2025_gprat} provides a highly parallel C\texttt{++} implementation of Gaussian process regression (GPR) with support for x86-64, ARM, and RISC-V platforms, leveraging the HPX runtime system~\cite{Kaiser2020}.
Unlike approaches that rely on third-party parallelization layers, GPRat explicitly manages parallelism at the application level.
To evaluate the effectiveness of this approach on emerging hardware architectures, we conduct a comprehensive performance study across three representative chips of a major transition phase in modern CPU design.
First, the AMD Zen 2 architecture introduces flexible chiplet-based designs, separating compute and I/O components into distinct dies.
Second, the Fujitsu A64FX chip exemplifies the successful adoption of ARM in HPC.
Finally, the SOPHON SG2042 chip represents a significant milestone for RISC-V and its arrival in HPC.

Three research questions guide this work: 1. Are contemporary chips based on the RISC-V architecture ready for ML workloads, and where are the areas that need improvement? 2. How does GPR with asynchronous tasks perform and scale across x86-64, ARM, and RISC-V? 3. Can OpenBLAS replace Intel oneMKL as a portable backend in GPRat?
Explicitly, our contributions in this work include:
\begin{itemize}
	\item The first evaluation of GPR for an ML application on the RISC-V architecture,
	\item The first comparison of GPR prediction and hyperparameter optimization on x86-64, ARM, and RISC-V,
	\item The extension of the GPRat library to support the ARM and RISC-V architectures,
	\item And a direct performance comparison of the Intel oneMKL and OpenBLAS libraries on the x86-64 architecture.
\end{itemize}

The remainder of this work is structured as follows.
Firstly, Section~\ref{sec:related_work} reviews related work regarding asynchronous many-task runtimes (AMTRs) and ML on RISC-V and ARM.
Secondly, in Section~\ref{sec:methods} we introduce the basic concepts of GPR.
Section~\ref{sec:framework} describes GPRat and the backends it relies on.
Subsequently, Section~\ref{sec:results} presents the results of the comparison between x86-64, ARM, and RISC-V.
Finally, in Section~\ref{sec:conclusion} we conclude our work and outline future work.

\section{Related Work}\label{sec:related_work}

In shared-memory computing environments, OpenMP~\cite{Dagum1998} is widely adopted due to its ease of use.
Despite its popularity, the fork-join execution model introduces unnecessary software-induced global synchronization points.
As an alternative, AMTRs have emerged, enabling more fine-grained parallelism.
Comprehensive overviews of existing AMTRs are provided by Thoman et al.~\cite{Thoman2018} and Schuchart et al.~\cite{Schuchart2025}.
Comparative performance studies on the x86-64 architecture are conducted using TaskBench~\cite{Slaughter2020_taskbench,Lahnor2026_taskbench_itoyori_hpx}.

In this work, we utilize the HPX runtime system~\cite{Kaiser2020}, which provides a C\texttt{++}-standard conform parallel programming model.
HPX supports ARM~\cite{Diehl2024_fugaku} and RISC-V~\cite{Diehl2023_riscv} architectures, as well as heterogeneous accelerators from NVIDIA, AMD, and Intel via Kokkos and SYCL integrations~\cite{Daiss2023_hpx_sycl,Daiss2022_hpx_kokkos}.
A growing ecosystem of applications has been developed on top of HPX. Notable examples include the astrophysical simulation code OctoTiger~\cite{Marcello2021_octotiger}, the distributed fast Fourier transform (FFT) tool HPX-FFT~\cite{Strack2024_hpxfft}, which also supports RISC-V platforms~\cite{Strack2026_riscv}, and the GPR library GPRat~\cite{Helmann2025_gprat}.

Several related works investigate the performance characteristics of RISC-V processors for HPC workloads.
Comparative evaluations between x86-64, ARM, and RISC-V architectures are reported in~\cite{Davik2024_riscv_arm,Mittone2023_riscv_arm_sg2042_energy,Simili2025_riscv_arm}.
The same RISC-V and ARM CPUs considered in this work are benchmarked for general HPC workloads in~\cite{Garade2026_riscv_arm} and for the OctoTiger application in~\cite{Diehl2023_riscv}.
The SOPHON SG2042 chip has received particular attention in recent performance studies.
Brown et al.~\cite{Brown2024_riscv_sg2042_isc,Brown2023_riscv_sg2042_sc} evaluate the CPU in comparison to contemporary x86-64 and ARM systems.
Venieri et al.~\cite{Venieri2026_riscv_cimonte} report roughly 30\% single-core GEMM performance gains for FP64 workloads using a specifically optimized OpenBLAS.
In contrast, Garcia et al.~\cite{Garcia2026_riscv_sg2042_llvm_rvv_openblas} show significant FP32 GEMM speedups by leveraging RISC-V vector extensions.
Beyond traditional HPC workloads, the SG2042 chip is also evaluated for artificial intelligence applications.
Neural network inference performance is investigated in~\cite{Malenza2025_riscv_ai}, while large language model (LLM) inference is studied in~\cite{Garcia2026_riscv_sg2042_llvm_rvv_openblas,Rodrigo2025_riscv_llm}.

Overall, existing work highlights the growing maturity of RISC-V and ARM platforms for HPC and ML workloads.
However, comprehensive evaluations of portable applications based on task-based parallelization remain limited.

\section{Gaussian Processes}\label{sec:methods}

GPs provide a flexible probabilistic framework for modeling complex nonlinear relationships.
The goal is to infer an unknown nonlinear function $f$ that maps input data from a feature matrix $Z = [\mathbf{z}_1, \mathbf{z}_2, \ldots, \mathbf{z}_n]^\top$, with $\mathbf{z}_i \in \mathbb{R}^D$, to a vector of observations $\mathbf{y} = [y_1, y_2, \ldots, y_n]^\top$, where $y_i \in \mathbb{R}$.
A GP is fully characterized by a mean function $m(\mathbf{z})$ and a covariance function, or kernel, $k(\mathbf{z}, \mathbf{z}')$.
For simplicity, the mean function is commonly assumed to be zero~\cite{Kocijan2016}, making the kernel function the primary determinant of the model's accuracy.
GPRat employs the squared exponential kernel:

\begin{equation}\label{eq:rbf}
	k\left(\mathbf{z}_i, \mathbf{z}_j \right) = \nu \cdot \exp \left( -\frac{1}{2\cdot l^2} \sum_{d=1}^{D} \left(z_{i,d} - z_{j,d} \right)^2 \right) + \delta_{ij} \sigma^2,
\end{equation}
where $D$ denotes the dimensionality of the input features.
The hyperparameters $l$, $\nu$, and $\sigma^2$ represent the length scale, signal variance, and noise variance, respectively.
The Kronecker delta $\delta_{ij}$ equals $1$ when $i=j$ and $0$ otherwise, accounting for observation noise on the training data.
The covariance matrix $K \in \mathbb{R}^{N \times N}$ can be constructed by evaluating all pairwise kernel values among the training samples.

Predictions for new unseen test samples $\hat{Z} \in \mathbb{R}^{M \times D}$ can be obtained from the posterior mean of the conditional distribution:

\begin{equation}\label{eq:pred}
	\hat{\mathbf{y}} = K_{\hat{Z}, Z} K^{-1} \mathbf{y},
\end{equation}
where $K_{\hat{Z}, Z} \in \mathbb{R}^{M \times N}$ denotes the cross-covariance matrix between training and test samples. The prior covariance matrix for the test data, $K_{\hat{Z}, \hat{Z}} \in \mathbb{R}^{M \times M}$, is computed analogously.
The posterior covariance matrix is given by

\begin{equation}\label{eq:uncert}
	\Sigma = K_{\hat{Z}, \hat{Z}} - K_{\hat{Z}, Z} K^{-1} K_{\hat{Z}, Z}^\top,
\end{equation}
where the diagonal elements of $\Sigma$ correspond to the predictive variances.
As the three hyperparameters of Equation (\ref{eq:rbf}) are the only tunable parameters of the entire model, a good hyperparameter selection is essential for accurate predictions.
This can be achieved by maximizing the marginal likelihood of the training data. For optimization purposes, the negative log marginal likelihood given by

\begin{equation}\label{eq:loglike}
	-\log \mathcal{L} (l, \nu, \sigma^2) = \frac{1}{2}\log |K| + \frac{1}{2} \mathbf{y}^\top K^{-1} \mathbf{y} + \frac{N}{2}\log (2 \pi).
\end{equation}
is minimized.
Optimization is performed using iterative methods such as gradient descent or the Adam optimizer~\cite{Kingma2015_adam}, by computing and updating the gradients of Equation (\ref{eq:loglike}) with respect to the hyperparameters.
The evaluation of Equations (\ref{eq:pred})–(\ref{eq:loglike}) requires the solution of linear systems involving the covariance matrix $K$.
Direct computation of the matrix inverse is generally avoided due to its numerical instability.
Instead, the symmetric positive-definite matrix $K$ is commonly factorized using the Cholesky decomposition.
Although this exact factorization has cubic computational complexity, once the lower-triangular Cholesky factor has been computed for use in Equations (\ref{eq:pred}) or (\ref{eq:loglike}), it can be reused efficiently to evaluate the predictive uncertainty and compute the log-determinant term.

\section{Software Framework}\label{sec:framework}

The main software tools that GPRat builds on are HPX for asynchronous parallelization and a BLAS/LAPACK backend, such as Intel MKL or OpenBLAS, for sequential but vectorized routines.

The GPRat library provides functions for prediction with or without uncertainty, hyperparameter optimization, loss computation, and utilities for managing the HPX runtime.
GPRat's core is its asynchronous tiled Cholesky decomposition.
The covariance matrix $K$ is divided into a set of tiles on which different LAPACK and BLAS kernels operate in a predetermined order~\cite{Helmann2025_gprat}. The tiled algorithm avoids redundant computations but leads to an inhomogeneous workload.
Parallelization of the tiled algorithms in GPRat is handled by the HPX runtime~\cite{Kaiser2020}.
A standout feature of HPX is its ability to execute code fully asynchronously via HPX futures, which allow for expressing task dependencies.
In addition to explicit future concatenation, the dataflow construct makes it easy to manage dependencies and build complex parallel task graphs.
With \texttt{hpx::dataflow}, GPRat can easily express the tiled algorithms required for the GP prediction and optimization pipeline.
Since GPRat explicitly parallelizes with asynchronous tasks, the chosen task size is a direct performance indicator and an additional external parameter to optimize.
Moreover, GPRat does not require a parallel BLAS library backend and can benefit from highly optimized sequential implementations.

Originally, GPRat was developed with support for Intel oneMKL, which provides highly optimized mathematical functions.
Since 2021, oneMKL has been part of the broader Intel oneAPI initiative.
Key performance optimizations include improved cache usage and vectorization up to 512-bit-wide AVX-512 on supported hardware.
The mathematical functions cover a wide range of BLAS and LAPACK routines, FFTs, sparse solvers, and more.
Specifically, the oneMKL library provides a complete implementation of the level one to three functions of the CBLAS interface standard, making it compatible with a wide variety of existing codebases.
In addition, oneMKL supports threading-based parallelization using OpenMP~\cite{Dagum1998} or Intel oneTBB~\cite{intel_tbb}.
Although oneMKL is tuned for Intel hardware, it also supports AMD x86-64 chips.
At the time of writing, there is no native ARM or RISC-V support.

To make GPRat portable, we add an OpenBLAS~\cite{Zhang2025_openblas} backend in this work.
Similar to the oneMKL library, the OpenBLAS library provides a set of optimized BLAS and LAPACK routines.
In contrast, OpenBLAS is open-source, not specifically optimized for Intel hardware, and does not include additional functions such as sparse solvers or FFTs.
OpenBLAS supports OpenMP~\cite{Dagum1998} and pthreads for threading-based parallelization.
In addition to x86-64 support, OpenBLAS also supports the ARM and RISC-V architectures.
On the Fujitsu A64FX chip, OpenBLAS partially supports 512-bit-wide Scalable Vector Extension (SVE).
In the OpenBLAS version we use in this work, the POTRF, TRSM, TRSV, and SYRK functions are not SVE-accelerated.
However, the functions that do the bulk of the work, such as GEMM and GEMV, are SVE-accelerated.
Regarding RISC-V, OpenBLAS is progressively adding vectorization support.
For the SOPHON SG2042, the RISC-V core vendor provides a special RVV 0.7.1 build chain~\cite{Venieri2026_riscv_cimonte}, whereas we use a more portable, generic OpenBLAS target in this work, which does not enable RVV.
Thus, in our generic build, the RISC-V chip runs scalar FP64 code, incurring a theoretical vectorization disadvantage of $4\times$ relative to x86-64 (AVX2, 256-bit, 4~doubles) and $8\times$ relative to ARM (SVE-512, 512-bit, 8~doubles).

Since OpenBLAS supports the CBLAS interface standard, we can use it as a drop-in replacement for oneMKL.
Due to its superior portability, OpenBLAS is the default backend since its introduction in GPRat v0.2.0.
The BLAS backend can be selected between oneMKL and OpenBLAS at compile time with the CMake variable \texttt{GPRAT\_ENABLE\_MKL}.
While GPRat was compared against GPyTorch~\cite{Gardner2018} and GPflow~\cite{Matthews2017} on x86-64 in~\cite{Helmann2025_gprat}, at the time of writing, these frameworks do not officially support RISC-V platforms. In contrast, GPRat builds natively with HPX and OpenBLAS on RISC-V, making it the appropriate choice for this cross-architecture comparison.

\section{Results}\label{sec:results}

\begin{table}[b]
	\centering
	\setlength{\tabcolsep}{8pt}
	\begin{tabular}{llll}
		\toprule
		Architecture & x86-64        & ARM           & RISC-V        \\ \midrule
		CPU          & AMD EPYC 7742 & Fujitsu A64FX & SOPHON SG2042 \\ 
		\rowcolor{lightgray!50}
		Cores        & 64            & 48            & 64            \\ 
		Base clock   & 2.25GHz       & 2.20GHz       & 2.00GHz       \\ 
		\rowcolor{lightgray!50}
		Process      & 7nm/14nm      & 7nm           & 12nm          \\ 
		L3 Cache     & 256MB         & n/a           & 64MB          \\ 
		\rowcolor{lightgray!50}
		RAM          & 2TB DDR4      & 32GB HBM2     & 128GB DDR4    \\ 
		SIMD         & AVX2          & SVE-512       & RVV 0.7.1     \\ 
		\rowcolor{lightgray!50}
		NUMA domains & 4             & 4             & 4             \\ 
		TDP          & 225W          & $\sim$140W    & 120W          \\ 
		\rowcolor{lightgray!50}
		Release      & 2019          & 2019          & 2023          \\ \bottomrule
	\end{tabular}
	\caption{Hardware specifications of the systems used for comparison.}
	\label{tab:spec}
\end{table}

Since GPs are widely employed in system identification~\cite{Kocijan2016}, we base our evaluation on data generated from a nonlinear mass-spring-damper simulator.
The simulator is included in the GPRat repository and enables the generation of datasets of arbitrary size.
The remainder of this section is structured into two parts.
First, we evaluate the hyperparameter optimization and prediction performance of GPRat across x86-64, ARM, and RISC-V platforms using a strong scaling benchmark.
In addition, we compare the performance of Intel oneMKL and OpenBLAS on the x86-64 architecture.
Second, we investigate the scalability of the considered ML workload with respect to problem size on all three hardware platforms.
All experiments are conducted using FP64 BLAS routines on a chip of each architecture.
Detailed hardware specifications are provided in Table~\ref{tab:spec}.
Runtimes presented in this section are averaged over ten runs, with 95\% confidence intervals shown as error bars.
Additional information regarding experimental reproducibility is provided in the~\nameref{sec:material} section.

\subsection{Strong Scaling}\label{sec:strong_scaling}

\begin{figure}
	\centering
	\begin{minipage}[t]{.47\textwidth}
		\centering
		\includegraphics[width=\linewidth]{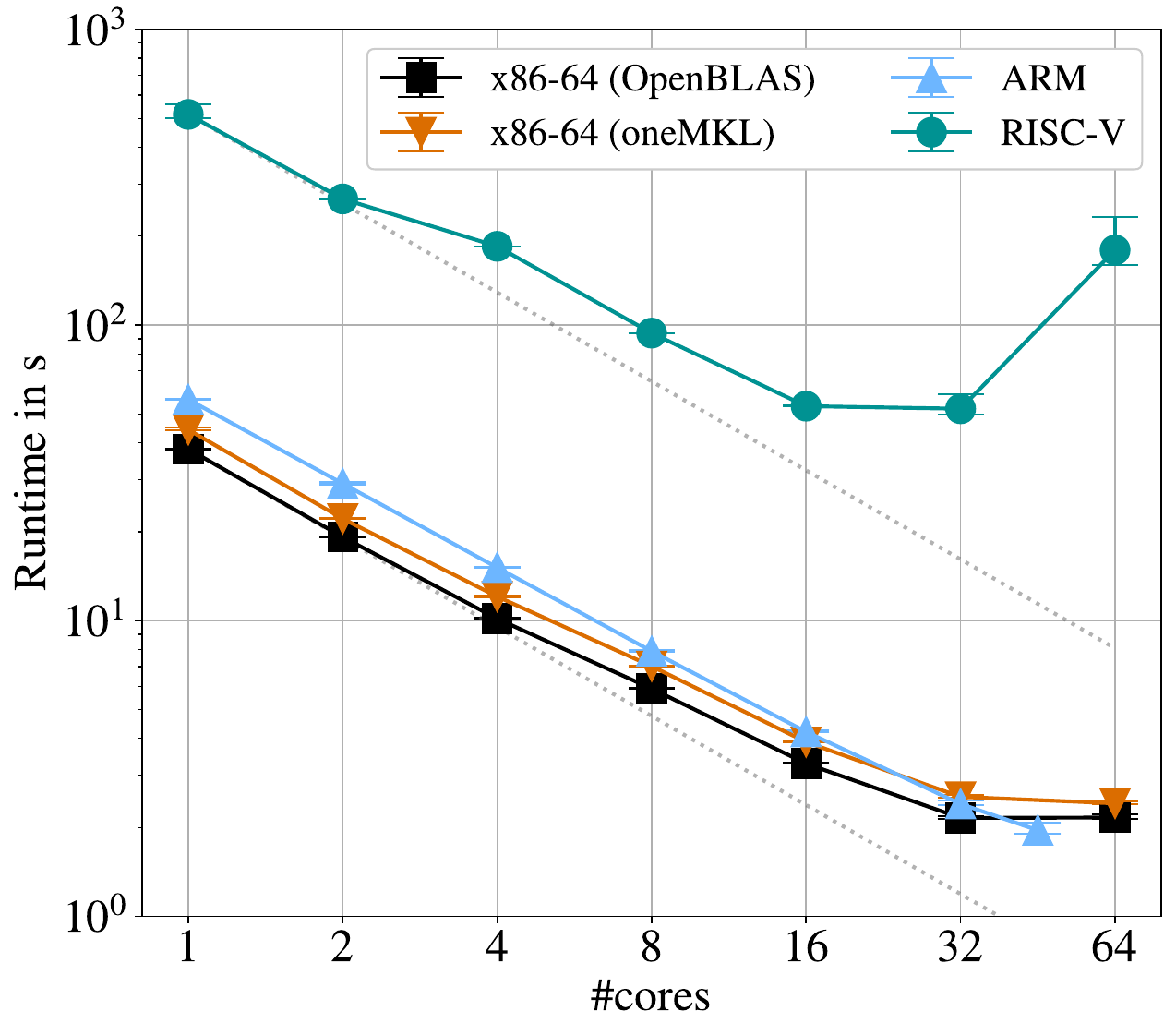}
		\caption{Strong scaling runtimes for hyperparameter optimization on up to $64$ x86-64, $48$ ARM, and $64$ RISC-V cores. The problem size was set to $N$$=$$2^{13}$ training samples. GPRat uses 16 tiles per dimension.}
		\label{fig:strong_opt_comp}
	\end{minipage}\hspace{.05\textwidth}
	\begin{minipage}[t]{.47\textwidth}
		\centering
		\includegraphics[width=\linewidth]{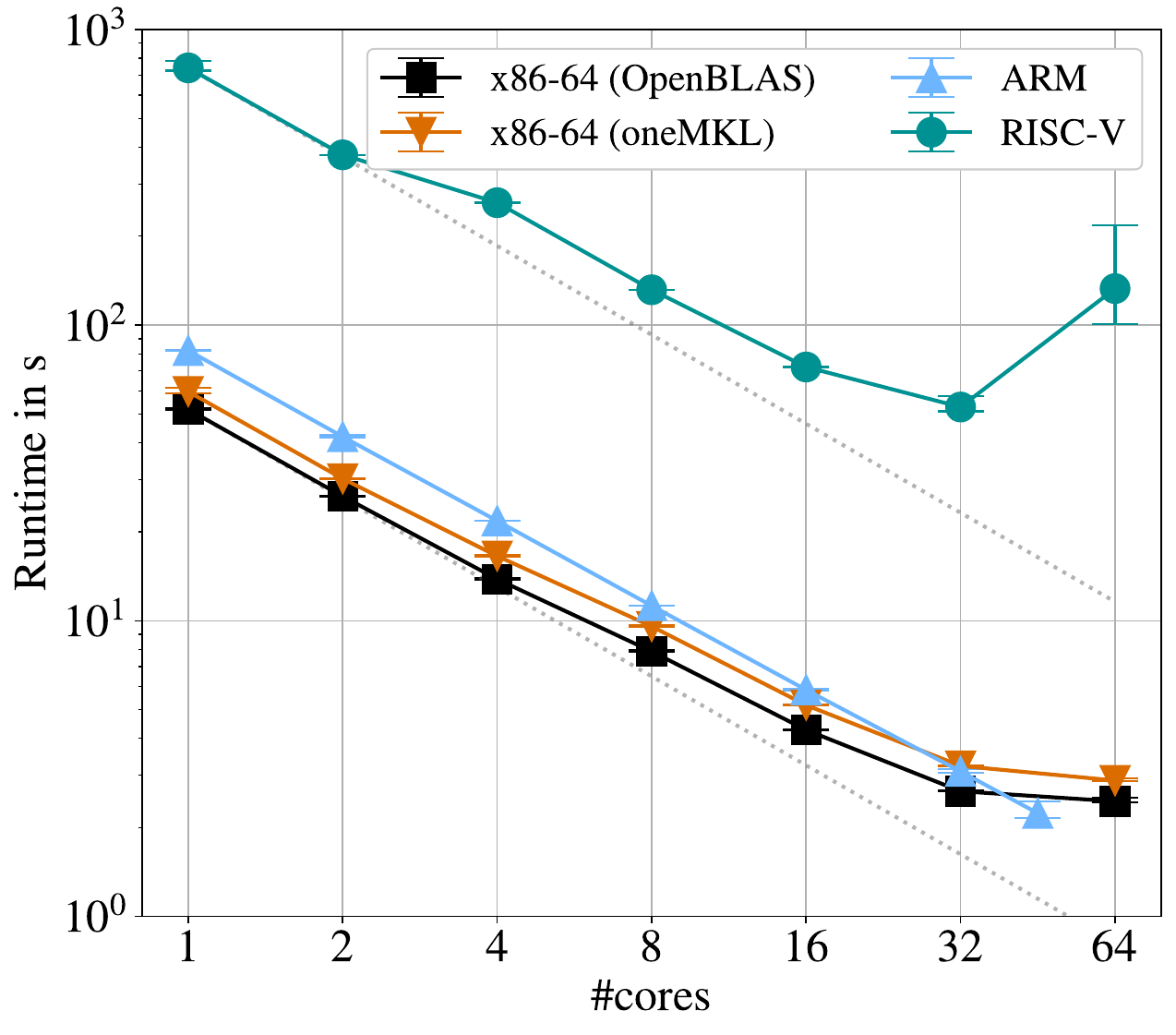}
		\caption{Strong scaling runtimes for prediction with full covariance matrix on up to $64$ x86-64, $48$ ARM, and $64$ RISC-V cores. The problem size was set to $N$$=$$M$$=$$2^{13}$ training samples and test samples. GPRat uses 16 tiles per dimension.}
		\label{fig:strong_pred_comp}
	\end{minipage}
\end{figure}

In Figures~\ref{fig:strong_opt_comp} and~\ref{fig:strong_pred_comp}, we compare the strong scaling of the oneMKL and OpenBLAS backends on x86-64 with the scaling of the OpenBLAS backends on ARM and RISC-V.
In direct comparison on the EPYC 7742 CPU, the OpenBLAS backend results in a speedup of $1.17$ for hyperparameter optimization and $1.19$ for prediction with full covariance.
The A64FX chip exhibits 58\% slower single-core performance but better scaling.
On 48 cores, the chip shows a 9\% performance improvement over the 64 Zen 2 cores while consuming significantly less power.
One reason for this behavior is a more efficient chip design and memory layout.
The ARM chip has no L3 cache but on-chip HBM2 memory.
Additionally, SVE-512 supports 512-bit vectorization for GEMM operations, unlike AVX2, which supports only 256-bit vectorization.
As GEMM is the most used routine in GPRat, SVE-512 GEMM acceleration is another factor contributing to the A64FX chip's good performance.

In contrast, on the RISC-V chip, scaling is heavily impaired.
These results are in line with the findings in~\cite{Brown2023_riscv_sg2042_sc} and~\cite{Strack2026_riscv}, where poor scaling was observed for 64 cores on the identical chip.
Furthermore, we observe a clear performance delta between RISC-V and the other two architectures.
Single-core performance is more than a factor of 14 slower than that of the x86-64 chips; correcting for the theoretical $4\times$ advantage of AVX2 over scalar code yields a scalar-to-scalar performance gap of approximately $3.5\times$, reflecting inherent architectural differences beyond vectorization.

To complement the runtime analysis, we report parallel efficiency.
ARM reaches $83\%$ / $73\%$ (prediction / optimization) efficiency at 32 cores; x86-64 with OpenBLAS achieves $61\%$ / $55\%$ efficiency at 32 cores but only $33\%$ / $28\%$ efficiency at 64 cores.
RISC-V begins at $70\%$ / $69\%$ efficiency at 8 cores but collapses to $9\%$ / $5\%$ efficiency at 64 cores.
These results confirm that scalability is impaired by the memory subsystem beyond the absence of vectorization and are consistent with~\cite{Brown2023_riscv_sg2042_sc,Strack2026_riscv}.

\subsection{Problem Size Scaling}\label{sec:problem_size}

In this subsection, we present the results of our problem size scaling experiments for hyperparameter optimization and prediction with full covariance for each hardware architecture.
The problem size, with the same number of training and test samples, is increased from $2^3$ up to $2^{14}$.

\begin{figure}
	\centering
	\begin{minipage}[t]{.47\textwidth}
		\centering
		\includegraphics[width=\linewidth]{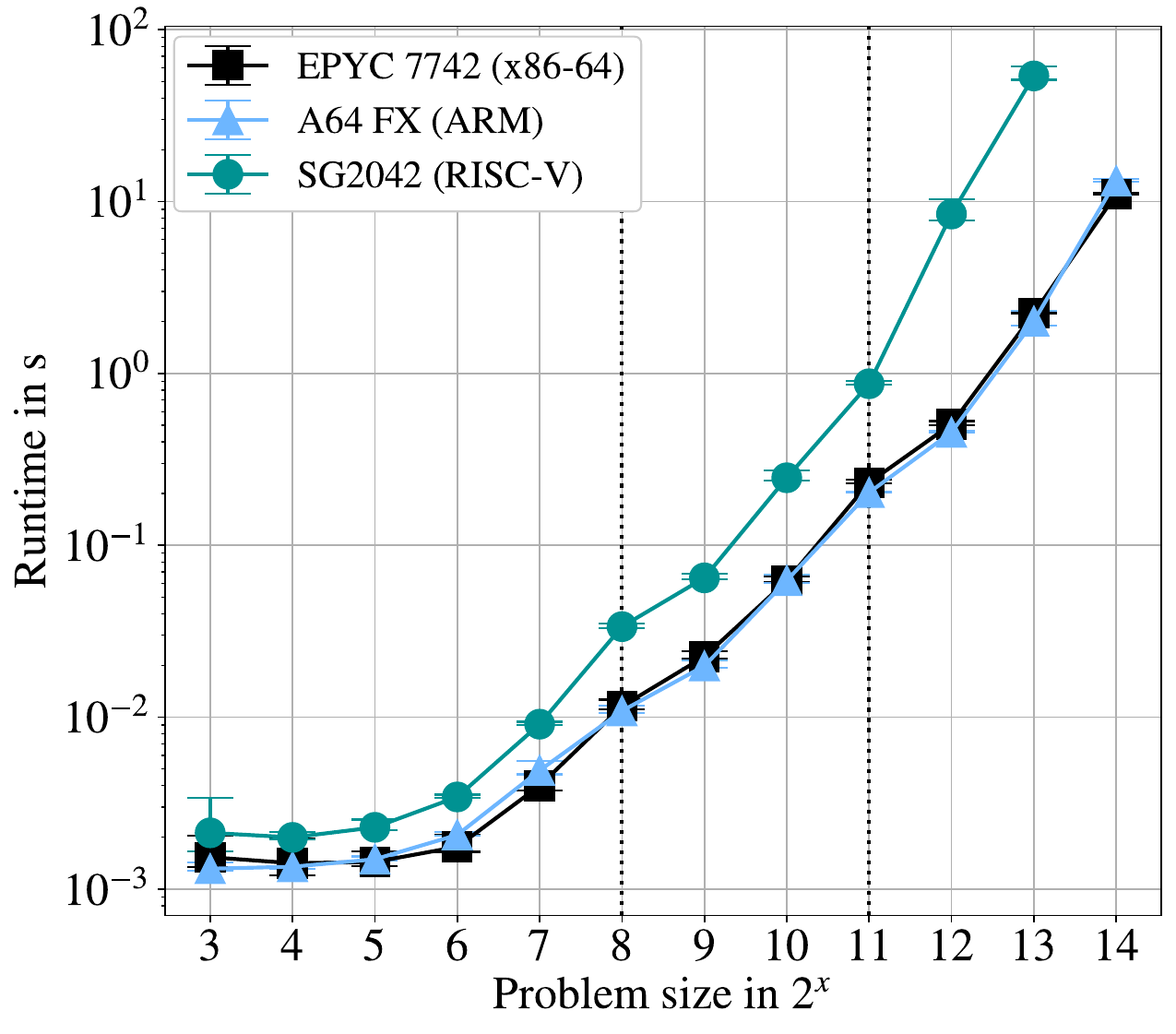}
		\caption{Problem size scaling runtimes for hyperparameter optimization on $64$ x86-64, $48$ ARM, and $32$ RISC-V cores.
			The number of tiles per dimension ranges from 1 to 16. Vertical dotted lines indicate changes.}
		\label{fig:problem_size_opt}
	\end{minipage}\hspace{.05\textwidth}
	\begin{minipage}[t]{.47\textwidth}
		\centering
		\includegraphics[width=\linewidth]{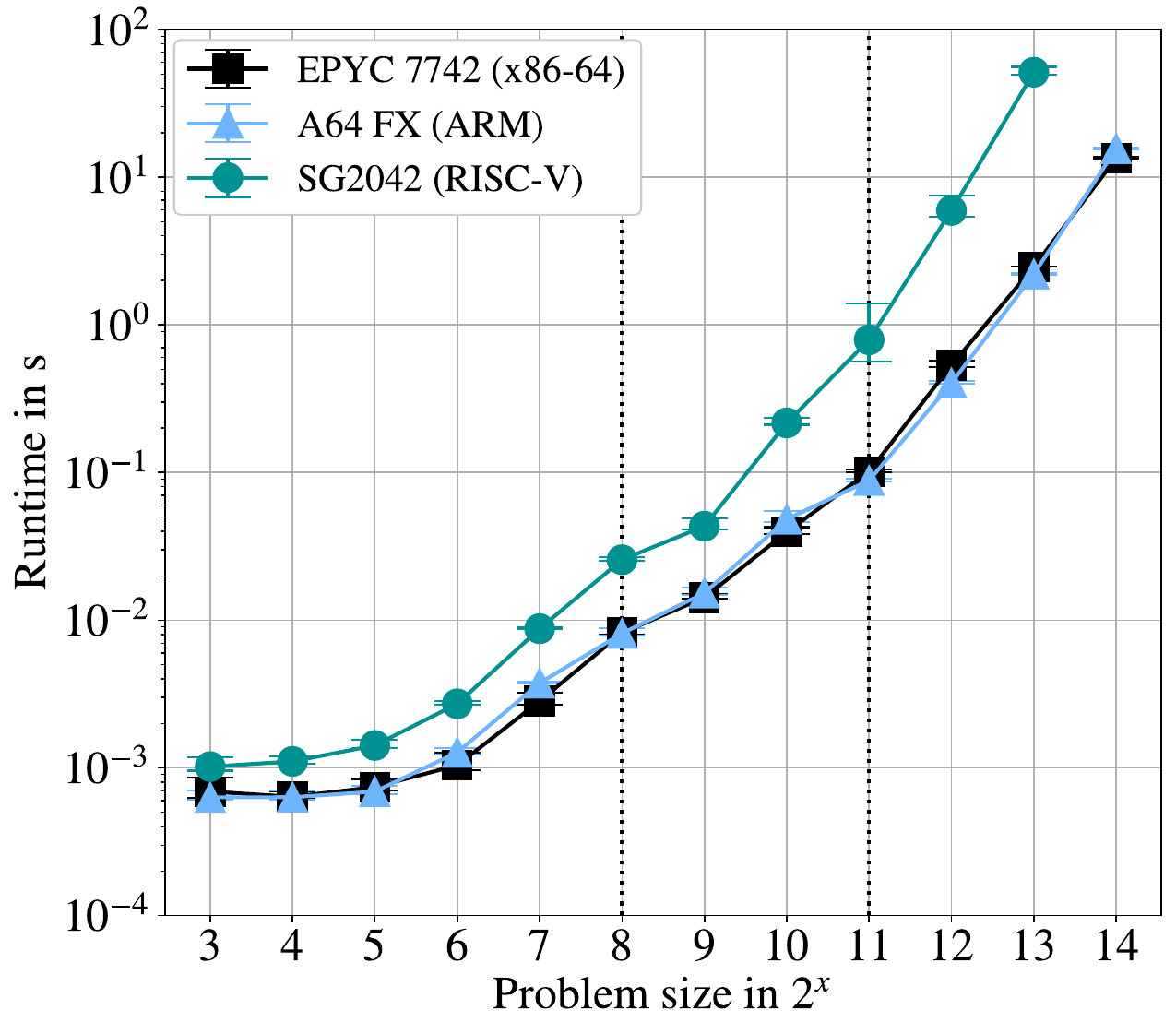}
		\caption{Problem size scaling runtimes for prediction with full covariance matrix on $64$ x86-64, $48$ ARM, and $32$ RISC-V cores.
			The number of tiles per dimension ranges from 1 to 16. Vertical dotted lines indicate changes.}
		\label{fig:problem_size_pred_full}
	\end{minipage}
\end{figure}

Figure~\ref{fig:problem_size_opt} shows the scaling runtimes for hyperparameter optimization across multiple problem sizes on all three chips.
We use different numbers of tiles per dimension, more specifically, one, four, and 16, indicated by vertical dotted lines.
Following~\cite{Helmann2025_gprat}, fewer tiles are optimal for small problem sizes, while more tiles improve performance for large sizes by exposing finer-grained parallelism.
We select the optimal core count for each chip: 64 for the x86-64 chip, 48 for the ARM chip, and 32 for the RISC-V chip.

The ARM and x86-64 chips exhibit similar performance within a 23\% performance delta.
For small sequential problems, the RISC-V has a performance delta of less than $3.1\times$ compared to the x86-64 chip.
For larger parallel problem sizes, however, the runtime delta increases up to a factor of 24.

Regarding the problem size scaling of the prediction with full covariance (see Figure~\ref{fig:problem_size_pred_full}), we observe similar behavior. For sequential problems, the performance delta ranges from 1.5 to 3.1, whereas for 4 and 16 tiles per dimension, the average performance deltas are up to 4.0 and 24.1, respectively.
The delta is exceptionally high for problem sizes of $2^{11}$ and larger.
These results highlight that while RISC-V single-core performance is getting more competitive, the lack of wide-register vectorization is a major bottleneck for HPC.
Furthermore, larger parallel workloads scale worse than on their ARM and x86-64 counterparts, which can be attributed to less efficient memory controller design.

\section{Conclusion and Outlook}\label{sec:conclusion}

In this work, we extended the GPR library GPRat to enable portability across emerging hardware architectures.
Furthermore, we conducted node-level strong scaling and problem size scaling benchmarks for prediction and hyperparameter optimization across three representative processors based on the x86-64, ARM, and RISC-V architectures.

In the strong scaling benchmark, we first compared Intel oneMKL and OpenBLAS on an EPYC 7742, where OpenBLAS achieved up to 19\% performance improvement.
With OpenBLAS, the x86-64 Zen 2 chip achieved a 58\% single-core performance advantage over the A64FX chip.
However, due to superior parallel scaling, the ARM chip ultimately outperformed the Zen 2 chip by 9\%.
In contrast, the RISC-V chip exhibited substantially lower single-core performance and inferior scaling behavior.

In the problem size scaling benchmark, the ARM and x86-64 chips demonstrate comparable performance, with differences within 23\%.
For small, sequential workloads, the performance gap between these architectures and the RISC-V chip narrows to less than a factor of three.
However, for large, parallel workloads, this gap widens substantially, reaching up to a factor of 24.

Overall, our results illustrate how emerging processor architectures handle compute-intensive GPR workloads dominated by dense linear algebra, findings that generalize broadly to other GEMM-heavy scientific computing and artificial intelligence applications.
Revisiting the research questions: For the chips tested in this work, we conclude that there is still a significant performance gap between RISC-V and both ARM and x86-64.
At the same time, reduced-instruction-set designs, such as the purpose-built A64FX chip, demonstrate that ARM-based architectures can compete effectively with contemporary state-of-the-art x86-64 processors.
The results also emphasize the importance of wide-register vectorization support and improvements to the memory subsystem in future RISC-V designs to achieve competitiveness at both the core and node levels with established HPC processors.
With respect to the BLAS backend, OpenBLAS outperforms oneMKL on x86-64 and ports seamlessly to ARM and RISC-V, making it a suitable drop-in replacement for oneMKL in GPRat.

Looking ahead, it will be informative to evaluate how next-generation ARM and RISC-V processors compare against competing x86-64 designs.
The successor to the SG2042, the SOPHON SG2044, has recently been evaluated in~\cite{Brown2025_riscv_sg2044}, demonstrating significantly improved memory bandwidth and scalability.
In addition, support for RVV 1.0 enables vectorization in generic builds.
On the ARM side, the A64FX will be succeeded by Fujitsu's Monaka chip, expected to be released in 2027.
Similar to recent AMD x86-64 chips, Monaka adopts a chiplet-based design that separates compute cores from I/O dies.
Another relevant factor to investigate is the performance per watt of all three architectures.

\begin{credits}
	\subsubsection{Supplementary Information}\label{sec:material}
	The GPRat library is open source under the MIT license and is available on \href{https://github.com/SC-SGS/GPRat/releases/tag/v0.2.0}{GitHub}\footnote{\url{https://github.com/SC-SGS/GPRat/releases/tag/v0.2.0} (Last accessed: 2026-05-26)}.
	Large data sets can be generated using the mass-spring-damper simulation in the GPRat repository.
	The data set we use in this work is available on \href{https://doi.org/10.18419/DARUS-4743}{DaRUS}\footnote{\url{https://doi.org/10.18419/DARUS-4743} (Last accessed: 2026-05-26)}.

	\subsubsection{\ackname} We gratefully acknowledge the LSU Center for Computation \& Technology for its support. The authors also thank Stony Brook Research Computing and Cyberinfrastructure, as well as the Institute for Advanced Computational Science at Stony Brook University, for providing access to the advanced high-performance computing system Ookami. The U.S. Department of Energy supported this work through the Los Alamos National Laboratory. Los Alamos National Laboratory is operated by Triad National Security, LLC, for the National Nuclear Security Administration of the U.S. Department of Energy (Contract No. 89233218CNA000001). Approved for public release with LA-UR-26-21448.

	\subsubsection{AI Usage Disclosure}

	Generative artificial intelligence (AI) tools, including Grammarly~\cite{grammarly}, ChatGPT~\cite{chatgpt}, and Claude~\cite{claude}, were employed to enhance the clarity, grammar, and overall coherence of the manuscript. All technical content, data analyses, and research findings were conceived and developed independently by the authors. AI-assisted outputs were carefully reviewed, verified, and edited by the authors to ensure factual accuracy, interpretive rigor, and scholarly integrity. The final manuscript reflects the authors' original intellectual contributions and analytical work.

\end{credits}
%
%
%
\bibliographystyle{splncs04}
\bibliography{main}

@book{Rasmussen2006,
  author    = {Carl Edward Rasmussen and Christopher K. I. Williams},
  title     = {Gaussian Processes for Machine Learning},
  publisher = {MIT Press},
  year      = {2006},
}

@article{Matthews2017,
  author  = {Matthews, Alexander G. de G. and {van der Wilk}, Mark and Nickson,
             Tom and others},
  title   = {{GP}flow: A {G}aussian process library using {T}ensor{F}low},
  journal = {JMLR},
  year    = {2017},
  volume  = {18},
  number  = {40},
  pages   = {1--6},
}

@inproceedings{Gardner2018,
  author    = {Gardner, Jacob R and Pleiss, Geoff and Bindel, David and others},
  title     = {Torch: Blackbox Matrix-Matrix {G}aussian Process Inference with
               {GPU} Acceleration},
  booktitle = {Adv. Neural Inf. Process. Syst.},
  year      = {2018},
}

@article{Kaiser2020,
  author  = {Hartmut Kaiser and Patrick Diehl and Adrian S. Lemoine and others},
  title   = {{HPX} -- The {C++} Standard Library for Parallelism and Concurrency},
  journal = {JOSS},
  year    = {2020},
  volume  = {5},
  number  = {53},
  pages   = {2352},
}

@book{Kocijan2016,
  author    = {J. Kocijan},
  title     = {Modelling and Control of Dynamic Systems Using {G}aussian Process
               Models},
  publisher = {Springer},
  year      = {2016},
}

@article{Dagum1998,
  author  = {Dagum, Leonardo and Menon, Ramesh},
  title   = {{OpenMP}: an industry standard {API} for shared-memory programming},
  journal = {IEEE Comput. Sci. Eng.},
  year    = {1998},
  volume  = {5},
  number  = {1},
  pages   = {46--55},
}

@article{Thoman2018,
  author  = {Thoman, Peter and others},
  title   = {A Taxonomy of Task-Based Parallel Programming Technologies for
             High-Performance Computing},
  journal = {J. Supercomput.},
  year    = {2018},
  volume  = {74},
  number  = {4},
  pages   = {1422--1434},
}

@techreport{Waterman2014_riscv,
  author      = {A. Waterman and others},
  title       = {{The RISC-V Instruction Set Manual, Volume I: User-Level ISA}},
  institution = {University of California, Berkeley},
  year        = {2014},
  number      = {2},
}

@manual{ARM2021,
  author = {{ARM Ltd.}},
  title  = {Arm Architecture Reference Manual {Armv8}, for {Armv8-A} architecture
            profile},
  year   = {2021},
  note   = {Available online},
}

@misc{Zhang2025_openblas,
  author       = {Xianyi Zhang and Qian Wang and Yunquan Zhang},
  title        = {{OpenBLAS}: An optimized {BLAS} library},
  howpublished = {\url{https://www.openblas.net}},
  year         = {2025},
}

@article{Diehl2024_fugaku,
  author  = {Diehl, Patrick and Daiß, Gregor and Huck, Kevin and others},
  title   = {Simulating stellar merger using {HPX/Kokkos} on {A64FX} on
             Supercomputer Fugaku},
  journal = {J. Supercomput.},
  year    = {2024},
  volume  = {80},
  pages   = {1--32},
}

@inproceedings{Diehl2023_riscv,
  author    = {Diehl, Patrick and others},
  title     = {Evaluating {HPX} and {Kokkos} on {RISC-V} using an astrophysics
               application {Octo-Tiger}},
  booktitle = {SC Workshops '23},
  year      = {2023},
  pages     = {1533--1542},
}

@article{Marcello2021_octotiger,
  author  = {Dominic C Marcello and Sagiv Shiber and Orsola De~Marco and others},
  title   = {{Octo-Tiger}: a new, {3D} hydrodynamic code for stellar mergers
             that uses {HPX} parallelization},
  journal = {Mon. Not. R. Astron. Soc.},
  year    = {2021},
  volume  = {504},
  number  = {4},
  pages   = {5345--5382},
}

@misc{Lahnor2026_taskbench_itoyori_hpx,
  author        = {Torben R. Lahnor and Mia Reitz and Jonas Posner and Patrick Diehl},
  title         = {Exploring Performance-Productivity Trade-offs in {AMT} Runtimes:
                   A Task Bench Study of Itoyori, {ItoyoriFBC}, {HPX}, and {MPI}},
  year          = {2026},
  eprint        = {2601.14608},
  archivePrefix = {arXiv},
  note          = {Preprint},
}

@inproceedings{Kingma2015_adam,
  author    = {Kingma, Diederik and Ba, Jimmy},
  title     = {Adam: A Method for Stochastic Optimization},
  booktitle = {ICLR},
  year      = {2015},
}

@misc{intel_tbb,
  author       = {{Intel Corporation}},
  title        = {{Intel oneAPI Threading Building Blocks (oneTBB)}},
  howpublished = {\url{https://www.intel.com/content/www/us/en/developer/tools/oneapi/onetbb.html}},
  year         = {2025},
  note         = {Version 2022.1},
}

@misc{Schuchart2025,
  author = {Schuchart, Joseph and Diehl, Patrick and Bauer, Michael and others},
  title  = {A Survey of Distributed Asynchronous Many-Task Models and Their
            Applications},
  year   = {2025},
  note   = {Preprint},
}

@inproceedings{Strack2026_riscv,
  author    = {Strack, Alexander and Taylor, Christopher and Pfl{\"u}ger, Dirk},
  title     = {{Parallel FFTW on RISC-V: A Comparative Study Including OpenMP,
               MPI, and HPX}},
  booktitle = {High Performance Computing},
  year      = {2026},
  publisher = {Springer},
  pages     = {586--597},
}

@inbook{Strack2024_hpxfft,
  author    = {Strack, Alexander and Taylor, Christopher and Diehl, Patrick and
               others},
  title     = {{Experiences Porting Shared and Distributed Applications to
               Asynchronous Tasks: A Multidimensional FFT Case-Study}},
  booktitle = {Asynchronous Many-Task Systems and Applications},
  publisher = {Springer},
  year      = {2024},
  pages     = {111--122},
}

@inproceedings{Daiss2023_hpx_sycl,
  author    = {Dai\ss{}, Gregor and Diehl, Patrick and Kaiser, Hartmut and others},
  title     = {{Stellar Mergers with HPX-Kokkos and SYCL: Methods of Using an
               Asynchronous Many-Task Runtime System with SYCL}},
  booktitle = {IWOCL '23},
  year      = {2023},
  publisher = {ACM},
}

@inproceedings{Helmann2025_gprat,
  author    = {Helmann, Maksim and Strack, Alexander and Pfl{\"u}ger, Dirk},
  title     = {{GPRat}: {G}aussian Process Regression with Asynchronous Tasks},
  booktitle = {Asynchronous Many-Task Systems and Applications},
  year      = {2026},
  publisher = {Springer},
  pages     = {83--94},
}

@inproceedings{Daiss2022_hpx_kokkos,
  author    = {Gregor Daiß and Srinivas Yadav Singanaboina and Patrick Diehl and
               others},
  title     = {From Merging Frameworks to Merging Stars: Experiences using {HPX},
               {Kokkos} and {SIMD} Types},
  booktitle = {ESPM2 '22},
  year      = {2022},
  publisher = {IEEE},
}

@inproceedings{Slaughter2020_taskbench,
  author    = {Slaughter, Elliott and Wu, Wei and Fu, Yuankun and others},
  title     = {Task Bench: A Parameterized Benchmark for Evaluating Parallel
               Runtime Performance},
  booktitle = {SC '20},
  year      = {2020},
  pages     = {1--15},
}

@inproceedings{Brown2025_riscv_sg2044,
  author    = {Brown, Nick},
  title     = {Is {RISC-V} ready for High Performance Computing? An evaluation
               of the Sophon {SG2044}},
  booktitle = {SC Workshops '25},
  year      = {2025},
  pages     = {1703--1711},
}

@inproceedings{Venieri2026_riscv_cimonte,
  author    = {Venieri, Emanuele and Manoni, Simone and Ceccolini, Gabriele and
               others},
  title     = {Monte {C}imone v2: {HPC} {RISC-V} Cluster Evaluation and
               Optimization},
  booktitle = {High Performance Computing},
  year      = {2026},
  publisher = {Springer},
  pages     = {576--585},
}

@inproceedings{Brown2023_riscv_sg2042_sc,
  author    = {Brown, Nick and Jamieson, Maurice and Lee, Joseph and others},
  title     = {Is {RISC-V} ready for {HPC} prime-time: Evaluating the 64-core
               Sophon {SG2042} {RISC-V} {CPU}},
  booktitle = {SC Workshops '23},
  year      = {2023},
  pages     = {1566--1574},
}

@inproceedings{Brown2024_riscv_sg2042_isc,
  author    = {Brown, Nick and Jamieson, Maurice},
  title     = {Performance Characterisation of the 64-Core {SG2042} {RISC-V}
               {CPU} for {HPC}},
  booktitle = {ISC High Perf. Workshops '24},
  year      = {2025},
  publisher = {Springer},
  pages     = {354--367},
}

@article{Davik2024_riscv_arm,
  author  = {Dak{\'i}{\'{c}}, Vedran and Mr{\v{s}}i{\'c}, Leo and Kuni{\'c},
             Zdravko and others},
  title   = {Evaluating {ARM} and {RISC-V} Architectures for High-Performance
             Computing with {Docker} and {Kubernetes}},
  journal = {Electronics},
  year    = {2024},
  volume  = {13},
  number  = {17},
  pages   = {3494},
}

@inproceedings{Garade2026_riscv_arm,
  author    = {Garade, Aniket P. and Bisht, Ashish and Deepika, H. V. and others},
  title     = {Evaluating {RISC-V} Processor as an Alternative for High
               Performance Computing},
  booktitle = {High Performance Computing},
  year      = {2026},
  publisher = {Springer},
  pages     = {521--533},
}

@inproceedings{Simili2025_riscv_arm,
  author    = {Emanuele Simili and Tommaso Boccali and Shahzad Muzaffar and others},
  title     = {Taking on {RISC} for Energy-Efficient Computing in {HEP}},
  booktitle = {EPJ Web of Conf.},
  volume    = {337},
  pages     = {01163},
  year      = {2025},
  publisher = {EDP Sciences},
  note      = {CHEP 2024},
}

@inproceedings{Rodrigo2025_riscv_llm,
  author    = {Poveda Rodrigo, Javier Jesus and Hamdi, Mohamed Amine and Koenig,
               Cyril and others},
  title     = {{V-Seek}: Optimizing {LLM} Reasoning on a Server-Class
               General-Purpose {RISC-V} Platform},
  booktitle = {CF '25},
  year      = {2025},
  publisher = {ACM},
  pages     = {224--225},
}

@article{Malenza2025_riscv_ai,
  author  = {Malenza, Giulio and Garcia, Adriano Marques and Birke, Robert and
             others},
  title   = {Analysis of Model Parallelism for {AI} Applications on a 64-core
             {RV64} Server {CPU}},
  journal = {Int. J. Parallel Program.},
  year    = {2025},
  volume  = {53},
  number  = {4},
  pages   = {27},
}

@inproceedings{Mittone2023_riscv_arm_sg2042_energy,
  author    = {Mittone, Gianluca and Tonci, Nicol\'{o} and Birke, Robert and others},
  title     = {Experimenting with Emerging {RISC-V} Systems for Decentralised
               Machine Learning},
  booktitle = {CF '23},
  year      = {2023},
  publisher = {ACM},
  pages     = {73--83},
}

@article{Garcia2026_riscv_sg2042_llvm_rvv_openblas,
  author  = {Adriano {Marques Garcia} and Giulio Malenza and Robert Birke and
             others},
  title   = {Inference performance of large language models on a 64-core {RISC-V}
             {CPU} with silicon-enabled vectors},
  journal = {Future Gener. Comput. Syst.},
  year    = {2026},
  volume  = {177},
  pages   = {108242},
}

@misc{chatgpt,
  author       = {{OpenAI}},
  title        = {{ChatGPT 5}},
  howpublished = {\url{https://openai.com/chatgpt}},
  year         = {2026},
  note         = {Last accessed: 2026-05-26},
}

@misc{grammarly,
  author       = {{Superhuman Platform Inc.}},
  title        = {Grammarly (Version 2.0)},
  howpublished = {\url{https://www.grammarly.com/}},
  year         = {2026},
  note         = {Last accessed: 2026-05-26},
}

@misc{claude,
  author       = {{Anthropic}},
  title        = {{Claude 4}},
  howpublished = {\url{https://claude.ai/}},
  year         = {2026},
  note         = {Last accessed: 2026-05-26},
}
\end{document}